\documentclass[prapplied,reprint,
superscriptaddress,
nofootinbib,
 amsmath,amssymb,
 aps,
floatfix,
]{revtex4-2}

\usepackage{amsmath}
\usepackage{graphicx}
\usepackage{dcolumn}
\usepackage{bm}

\usepackage[utf8]{inputenc}
\usepackage[T1]{fontenc}
\usepackage{mathptmx}
\usepackage{etoolbox}
\usepackage{siunitx}

\def\reals{{{\rm l} \kern-.15em {\rm R}}}

\newcommand{\xb}{{\bf x}}
\newcommand{\yb}{{\bf y}}

\newcommand{\CCb}{{\bf \mathcal{C}}}

\usepackage{xcolor}

\newcommand{\comment}[1]{\textcolor{red}{}}

\makeatletter
\def\@email#1#2{%
 \endgroup
 \patchcmd{\titleblock@produce}
  {\frontmatter@RRAPformat}
  {\frontmatter@RRAPformat{\produce@RRAP{*#1\href{mailto:#2}{#2}}}\frontmatter@RRAPformat}
  {}{}
}%
\makeatother
\begin{document}

\preprint{AIP/123-QED}

\title[Precise 2D electric field density simulations for superconducting quantum devices]{Precise 2D electric field density simulations for superconducting quantum devices}
\author{Z.~Gimbutas}
\affiliation{National Institute of Standards and Technology Boulder}
\author{W.-R.~Syong}
\affiliation{Department of Physics, University of Colorado Boulder}
\affiliation{Department of Electrical, Computer, and Energy Engineering, University of Colorado Boulder}
\affiliation{National Institute of Standards and Technology Boulder}
\author{N.~Nguyen}
\affiliation{Department of Physics, University of Colorado Boulder}
\affiliation{Department of Electrical, Computer, and Energy Engineering, University of Colorado Boulder}
\affiliation{National Institute of Standards and Technology Boulder}
\author{A. Taylor}
\affiliation{Department of Physics and Astronomy, University of Texas at San Antonio}
\author{J.~N.~Ullom}%
\affiliation{National Institute of Standards and Technology Boulder}
\affiliation{Department of Physics, University of Colorado Boulder}
\author{B.~K.~Alpert}%
\affiliation{National Institute of Standards and Technology Boulder}
\author{D.~A.~Bennett}%
\affiliation{National Institute of Standards and Technology Boulder}
\affiliation{Department of Electrical, Computer, and Energy Engineering, University of Colorado Boulder}
\author{C.~R.~H.~McRae}
\affiliation{Department of Electrical, Computer, and Energy Engineering, University of Colorado Boulder}
\affiliation{Department of Physics, University of Colorado Boulder}
\affiliation{National Institute of Standards and Technology Boulder}

\date{\today}

\begin{abstract}
Dielectric loss due to two-level systems is a limiting factor for superconducting qubit relaxation times. These losses arise mostly from nanometer-scale interfacial defect regions in superconducting devices with planar dimensions of microns to millimeters, thus making it resource intensive to accurately simulate the electric field density in these regions with traditional electromagnetic solvers. In this work, we demonstrate a fast boundary integral equation solver that allows precise simulation of electric field density in these thin regions, showing a speedup of around two orders of magnitude over traditional solvers, with relative errors around $10^{-7}$ for a ten-minute solution runtime. By computing participation ratios through Green's first identity without squaring the electric field, our approach is less susceptible to the field singularities near conductor corners. We apply this solver to a basic untrenched coplanar waveguide cross-section, showing that the common assumption of participation ratio linearity with dielectric constant holds well for some interfaces and not others; in particular, while the metal-air (MA) top and corner follow this linear relationship strongly, the MA sidewall does not. We then compare isotropic and anisotropic etching, showing that the MA sidewall and the metal-air-substrate triple junction are the most strongly affected. We are currently leveraging this solver to explore geometries that will uniquely isolate the participation ratios of the different dielectrics. Finally, we are working to combine this solver framework with a full 3D microwave solver to accurately calculate participation ratios for the thin dielectrics that are known sources of loss in superconducting qubits.

\end{abstract}

\maketitle

\section{\label{sec:level1}Introduction}

For decades, superconducting qubits have been performance-limited by dielectric loss arising from two-level systems~\cite{Martinis2005,Muller2019}. These losses arise largely from thin (nanometer-scale) interfaces between the superconducting thin film, the dielectric substrate, and the air or vacuum surrounding the device. 

In order to disentangle loss contributions from each of these regions and thus understand and ultimately improve superconducting qubit performance, the participation ratio obtained from the electric field density must be known for each region~\cite{Wenner2011,Wang2015,Calusine2018,Woods2019,McRae2020}. This requirement creates a challenge for solvers, since some regions with small participation ratios---of a few parts per million---have very large loss tangents and therefore large contributions to loss.  The solver therefore needs to be able to estimate participation ratios with considerably better than ppm accuracy.  

\begin{figure}
\includegraphics[width=\linewidth]{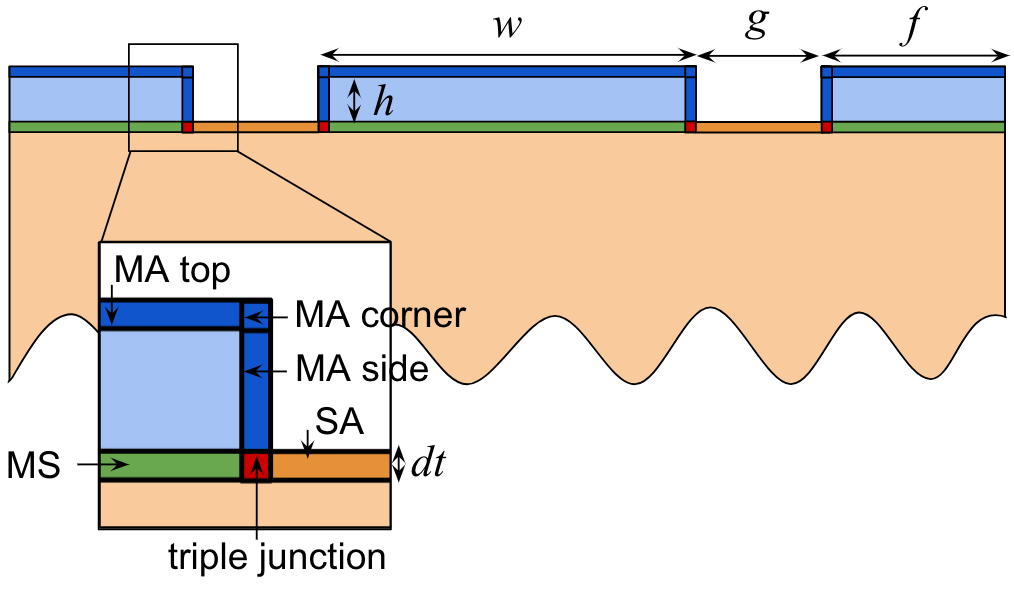}
\includegraphics[width=\linewidth]{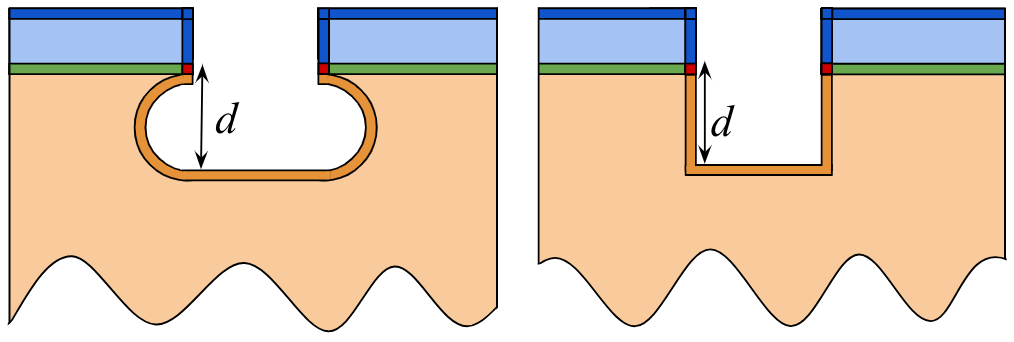}
\caption{\label{fig:CPWdiagram}
Cross-sectional diagram of coplanar waveguide (CPW). The CPW structure is defined by conductor width $w$, gap width $g$, superconducting metal thickness $h$, ground plane width $f$, and trench depth $d$. Trenching can have an isotropic (bottom left) or anisotropic (bottom right) profile. Interfaces are labeled as metal-substrate (MS, green), substrate-air (SA, dark orange), metal-air (MA, dark blue), and triple junction (TJ, red), and are defined by thickness $dt$.
}
\end{figure}

Coplanar waveguide (CPW) resonators are often used as proxy devices to identify losses in superconducting qubits~\cite{McRae2020}, and thus it is of particular interest to identify the participation ratios for interfacial regions in these devices. Conformal mapping can be used for some CPW geometries to identify exact solutions~\cite{Simons2001,Murray2018}, but these require significant simplifications such as infinitely thin metal layers and dielectric interfaces. 

Electromagnetic simulations are commonly used to capture more realistic device geometries. These simulations are not trivial, as the difference in critical dimensions is around three orders of magnitude -- for example, the CPW center conductor width and gaps are typically 3--10 $\mu$m, while the oxide layers are only 1--3 nm thick. CPWs are less costly to simulate than qubits, as their quasi-transverse-electromagnetic (quasi-TEM) nature allows us to perform an electrostatic simulation of a representative 2D cross-section, avoiding 3D high-frequency simulations~\cite{dial2016bulk_surf,Gambetta2017,Niepce2020}. Solver requirements include the ability to solve for participation ratios in various regions in a model with at least two conductors at different potentials, as well as multiple thin dielectric layers.

Commonly-used electrostatic solvers such as COMSOL~\cite{Calusine2018,Melville2020,Woods2019,Lahtinen2020} and Ansys Maxwell~\cite{Crowley2023} use finite elements to mesh the 2D model. Divergent electric fields at corner regions require an extremely fine mesh to obtain a converging participation ratio solution, leading to long simulation times and large memory requirements~\cite{Wenner2011} for these current methods.

Because of the resource-intensive nature of these simulations, it is common practice to rely on assumptions of participation linearity with respect to interface dielectric constant and thickness~\cite{Calusine2018,Melville2020,Woods2019}, which have not been probed extensively but do not seem to hold generally~\cite{Lahtinen2020}. The ability to perform more rapid simulations of this type would allow a better understanding of CPW participation ratios and their dependence on multiple variables such as sidewall parameters, trenching, and dielectric constant. Additionally, to the authors' knowledge, there have been no previous attempts to perform an error analysis on the extracted values from these simulations.

In this study, we demonstrate a fast solver from a boundary integral equation formulation that allows precise simulation of electric field density in thin dielectric regions, showing a speedup of around two orders of magnitude over Ansys Maxwell, with relative errors around $10^{-7}$ for a solution with a ten-minute runtime. We then apply this solver to CPW cross-sectional geometries, showing a deviation from conventional dielectric constant linearity assumptions for the metal-air (MA) sidewall participation but not for the MA top. Finally, we apply the solver to evaluate how the participation ratios in the MA sidewall are impacted by trenching of the substrate.

\section{Background information}

The participation ratio is the fraction of the total electric field energy stored in a given region of a device. For dielectric region $\Omega_i$, the electric field energy per unit length stored in that region is
\begin{equation}
W_i = \int_{\Omega_i} \frac{\varepsilon_i}{2} |E|^2\,dS,
\end{equation}
where $\varepsilon_i$ is the permittivity, $E$ is the electric field, and $dS$ is the cross-sectional area element. The participation ratio is then defined as
\begin{equation}
p_i = \frac{W_i}{W} = \frac{\int_{\Omega_i} \varepsilon_i
|E|^2\,dS}{\sum_j\int_{\Omega_j} \varepsilon_j |E|^2\,dS}.
\end{equation}

For the case of a 2D CPW cross-section, it is common to identify four main regions of interest, as shown in Fig.~\ref{fig:CPWdiagram}: the metal-air (MA), substrate-air (SA), and metal-substrate (MS) interfaces, and the substrate (sub). Ideally, these regions are well-defined, homogeneous, and have distinct dielectric constants and thicknesses. The ability to distinguish loss from these regions allows the identification of limiting factors in superconducting qubit performance. 

In this work, we further subdivide the MA region into MA top, MA side, and MA corner. Additionally, we label the square sitting between the MA, MS, and SA regions as the triple junction (TJ). The two-level system (TLS) loss tangent of the CPW resonator is then
\begin{align}
    \tan\delta_{TLS} &= 
    p_{\mathrm{MS}}\tan\delta_{\mathrm{MS}} + p_{\mathrm{SA}}\tan\delta_{\mathrm{SA}}\nonumber
    \\& + (p_{\mathrm{MA,side}} + p_{\mathrm{MA,top}} + p_{\mathrm{MA,corner}})\tan\delta_{\mathrm{MA}}\nonumber
    \\& + p_{\mathrm{TJ}}\tan\delta_{\mathrm{TJ}} + p_{\mathrm{sub}}\tan\delta_{\mathrm{sub}},
\end{align}
where the loss tangent $\tan\delta$ is the ratio of the imaginary to the real part of the permitivity.
\section{Solver}

Our solver is a 2D electrostatic capacitance solver for multi-conductor transmission line cross-sections, based on a boundary integral equation (BIE) method. The capacitance matrix is obtained by applying unit voltage excitations to each conductor independently, integrating the resulting surface charges, and applying a charge-neutrality correction~\cite{paulfeather1976capmat}. The underlying boundary value problem is: given piecewise-constant permittivities $\varepsilon(\xb)$ and prescribed conductor potentials $V_i$, find the scalar potential $\phi$ satisfying
\begin{equation}\label{eq:bvp}
  \begin{split}
    \nabla\cdot \varepsilon \nabla \phi &= 0, \quad \text{in each homogeneous region,}\\
    \phi &= V_i, \quad \text{on conductor $i$,}\\
    [\phi] &= 0, \quad \text{on dielectric interfaces,}\\
    \left[\varepsilon\, \frac{\partial\phi}{\partial n}\right] &= 0, \quad \text{on dielectric interfaces,}\\
    \phi &\to 0, \quad \text{as $|\xb| \to \infty$.}
  \end{split}
\end{equation}
where $[\cdot]$ denotes the jump across an interface and $n$ is the outward unit normal.

We express~$\phi$ as the potential due to a charge density~$\sigma$ distributed on the boundaries (a single layer potential representation), reducing the problem to a boundary integral equation for~$\sigma$. The boundary values of~$\phi$ and $\partial\phi/\partial n$ are then recovered from~$\sigma$ by evaluating the layer potential and its normal derivative.

In addition to the capacitance matrix, the solver computes the participation ratios for each dielectric region. Near conductor corners, the electric field diverges as $|E| \sim r^{-\alpha}$ (where $r$ is the distance from the corner and $\alpha$ is the angle-dependent exponent), but the physical edge condition requires the energy density $\varepsilon |E|^2$ to remain integrable in the dielectric regions adjacent to the corners~\cite{bouwkamp1946,meixner1949,meixner1972}. The participation ratios are computed directly from $\phi$ and $\partial\phi/\partial n$ via Green's first identity~\cite{strauss2008partial}:
\begin{equation}\label{eq:greenid}
  p_i = W^{-1} \int_{\partial \Omega_i} \frac{\varepsilon_i}{2}\, \phi\, \frac{\partial \phi}{\partial n}\, ds,
\end{equation}
where $W$ is the total electric field energy per unit length and $\partial \Omega_i$ is the boundary of dielectric region $\Omega_i$. This identity converts the area integral of $\varepsilon|E|^2$ into a boundary integral involving the product $\phi\,\partial\phi/\partial n$, where $\partial\phi/\partial n$ is singular but integrable near corners, while $\phi$ remains bounded --- avoiding squaring the singular field. 

As an independent error check, we verify participation ratios using a perturbative method~\cite{schneider1969microstrip,spielman1977dissipation}: perturbing each region's dielectric constant and computing the resulting change in capacitance (see Appendix~A). We also monitor $\sum_i p_i$, which must equal unity, as a global consistency check.

The solver handles arbitrary multi-conductor and multi-dielectric configurations, including microstrips and coplanar waveguides with isotropic and anisotropic trench profiles, arbitrary oxide layers, and flip-chip architectures. Cross-section geometries are defined by parameterized descriptors covering the configurations listed above; the integral equation formulation and implementation details are given in Appendices~A and~B.  

We validate the solver against exact solutions and established commercial solvers. For a pair of circular wires, we obtain 13 digits of accuracy in the capacitance; for a square coaxial waveguide with corner singularities similar to those in CPW geometries, we achieve ten digits of accuracy (see Appendix~C for details).

We compare our solver against Ansys Maxwell for a CPW cross-section, varying the ground plane width~$f$ to study convergence. Our solver achieves relative errors below 1\% for all participation ratios at $f > 20$~$\mu$m, while Ansys Maxwell requires $f > 40$~$\mu$m. All participation ratios agree to within 1\% between the two solvers, except for $p_{\text{MA,corner}}$ (Table~\ref{tab:maxwellcomp_v4}), and our solver is faster by a factor of approximately 170. Our results also reproduce the COMSOL simulations in Wenner et al.~\cite{Wenner2011} to within fractions of a percent (Table~\ref{tab:wennercomparison}). Full comparison details are provided in Appendices~D and~E.

\section{Dielectric constant dependence}

We now focus on participation ratios for 2D cross-sections of a coplanar waveguide (CPW) resonator. CPW cross-section geometry is shown in Fig.~\ref{fig:CPWdiagram} for an untrenched geometry (top), an isotropically trenched geometry (bottom left), and an anisotropically trenched geometry (bottom right). The CPW has conductor width $w$ and gap width $g$. A finite superconducting metal thickness $h$ is used. We define metal regions as perfect superconductors. The ground plane width $f$ defines the domain size in the x direction. The substrate is semi-infinite and has dielectric constant $\varepsilon_{\mathrm{sub}}$. The etched geometries have trench depth $d$. The height of air and substrate in the y direction is not truncated in this simulation. Thin dielectric interfaces are broken into six main regions, all with characteristic thickness $dt$.

Although it is common practice to assume that participation has a linear (or inversely linear) relationship to dielectric constant, there is evidence to suggest that this assumption does not always hold~\cite{Calusine2018,Woods2019,Lahtinen2020}. Here, we take advantage of the speed and accuracy of our solver to perform a detailed analysis of the electric field linearity and participation of each CPW region as a function of interface dielectric constants, and further detail when this common assumption holds and when it does not.

\subsection{Linearity assumptions}

For very thin interfacial regions, we can further break down the participation ratio into parallel and perpendicular components,~\cite{Woods2019}
\begin{align}
p_{i} &= \frac{\int_{\Omega_i}\varepsilon_i|E_{\parallel}|^2\,dS}{\sum_j\int_{\Omega_j}\varepsilon_j|E|^2\,dS} + \frac{\int_{\Omega_i}\varepsilon_i|E_{\perp}|^2\,dS}{\sum_j\int_{\Omega_j}\varepsilon_j|E|^2\,dS},\nonumber
\\
&= p_{i,\parallel} + p_{i,\perp}
\end{align}
and assume that each interface has exclusively a parallel or perpendicular component, as follows:
\begin{align}
p_{\mathrm{MS}} &\sim p_{\mathrm{MS},\perp}\nonumber
\\
p_{\mathrm{MA}} &\sim p_{\mathrm{MA},\perp}\label{surfacePR}
\\
p_{\mathrm{SA}} &\sim p_{\mathrm{SA},\parallel}.\nonumber
\end{align}

Assuming that the energy density in each interface is independent of that in the other regions, we can reasonably assume that the participation scales linearly or inversely linearly with both region thickness $t$ and dielectric constant $\varepsilon$. So, given a simulated participation ratio $p_{i,\parallel}^{\mathrm{sim}}$ or $p_{i,\perp}^{\mathrm{sim}}$ where dielectric constant is set to $\varepsilon_{\mathrm{sim},i}$ and thickness is set to $t_{\mathrm{sim},i}$, we can determine the participation for any thickness $t_i$ and dielectric constant $\varepsilon_i$,
\begin{align}
p_{i,\parallel} &= \frac{t_i/t_{\mathrm{sim},i}}{\varepsilon_{\mathrm{sim},i}/\varepsilon_i}p_{i,\parallel}^{\mathrm{sim}}\nonumber
\\
p_{i,\perp} &= \frac{t_i/t_{\mathrm{sim},i}}{\varepsilon_{i}/\varepsilon_{\mathrm{sim},i}}p_{i,\perp}^{\mathrm{sim}}.
\end{align}

The above is a critical assumption for the analysis of experimental devices, as it allows the identification of interface participation ratios with unknown thicknesses and dielectric constants. This is especially crucial for regions such as the metal-substrate interface, which in practice is not usually a well-defined region with a distinct dielectric constant.

Applying the above to the MS, SA, and MA regions, we see:~\cite{Lahtinen2020}
\begin{align}
p_{\mathrm{SA}} &= (\varepsilon_{\mathrm{SA}}/\varepsilon^{\mathrm{sim}}_{\mathrm{SA}}) \, p^{\mathrm{sim}}_{\mathrm{SA}}\nonumber
\\
p_{\mathrm{MA}} &= (\varepsilon^{\mathrm{sim}}_{\mathrm{MA}}/\varepsilon_{\mathrm{MA}}) \, p^{\mathrm{sim}}_{\mathrm{MA}}
\label{eq:dielconstlin}\\
p_{\mathrm{MS}} &= (\varepsilon^{\mathrm{sim}}_{\mathrm{MS}}/\varepsilon_{\mathrm{MS}}) \, p^{\mathrm{sim}}_{\mathrm{MS}},\nonumber
\end{align}
with the implication that each participation ratio is independent of the dielectric constants in all other regions.

Our simulation to test this assumption is a CPW cross-section with $w$ = 6~$\mu$m, $g$ = 3~$\mu$m, $dt$ = 3~nm, $\varepsilon_{\mathrm{sub}}$ = 10, $h$ = 100~nm, and $f$ = 300~$\mu$m. We compare an untrenched ($d$ = 0~nm) and an anisotropically-trenched ($d$ = 20~nm) version of this cross-section.

\subsection{Linearity tests}

\begin{figure}
\includegraphics[width=1.0\linewidth]{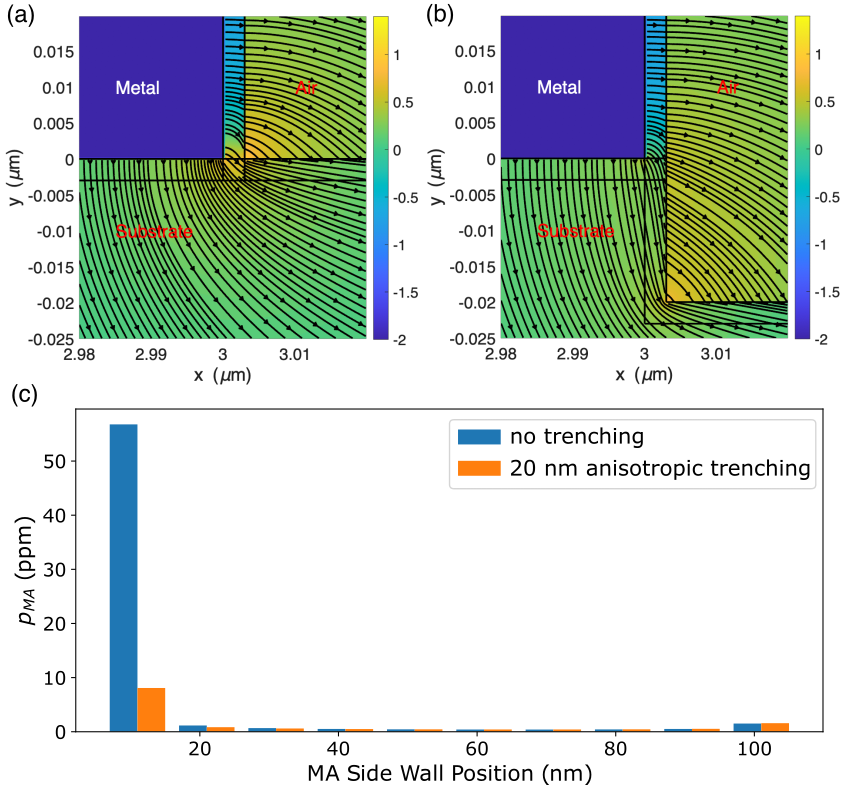}
\caption{\label{fig:sidewalldirection}
  Electric field density (colormap) and electric field direction (streamlines) near the bottom of the CPW center conductor sidewall for (a) an untrenched and (b) a 20 nm anisotropically trenched geometry. (c) The participation ratio for each of the 10 subdivided MA side subregions, starting from the bottom of the MA sidewall.
}
\end{figure}

Fig.~\ref{fig:sidewalldirection}(a) and (b) show the directionality of the electric field in the interfaces near the TJ in both an untrenched and anisotropically-trenched CPW. Away from the TJ, the field propagates almost perpendicularly or in parallel with the interface, as described in the above assumptions. However, close to the TJ, the field has significant components in both directions.

By further breaking down the MA side region into subregions as shown in Fig.~\ref{fig:sidewalldirection}(c), it is evident that the MA sidewall participation is dominated by the nonlinear region closest to the TJ for both trenched and untrenched geometries. However, the dominance is significantly more stark in the untrenched geometry.

Next, we individually sweep interface dielectric constants $\varepsilon_{\mathrm{MA}}$, $\varepsilon_{\mathrm{MS}}$, $\varepsilon_{\mathrm{SA}}$, and $\varepsilon_{\mathrm{TJ}}$ between 1 and 50, keeping all other regions at a dielectric constant of 10. Then, we compare the participation ratios to the relationships above. Results for every region are shown in the supplemental, including the MS, SA, MA top, and MA corner regions that closely follow the expected linear relationships. These four regions can be confidently analyzed using the assumptions in Eq.~\ref{eq:dielconstlin} that are the basis of the loss factor approach~\cite{Woods2019}.

\begin{figure}[t]
\includegraphics[width=0.9\linewidth]{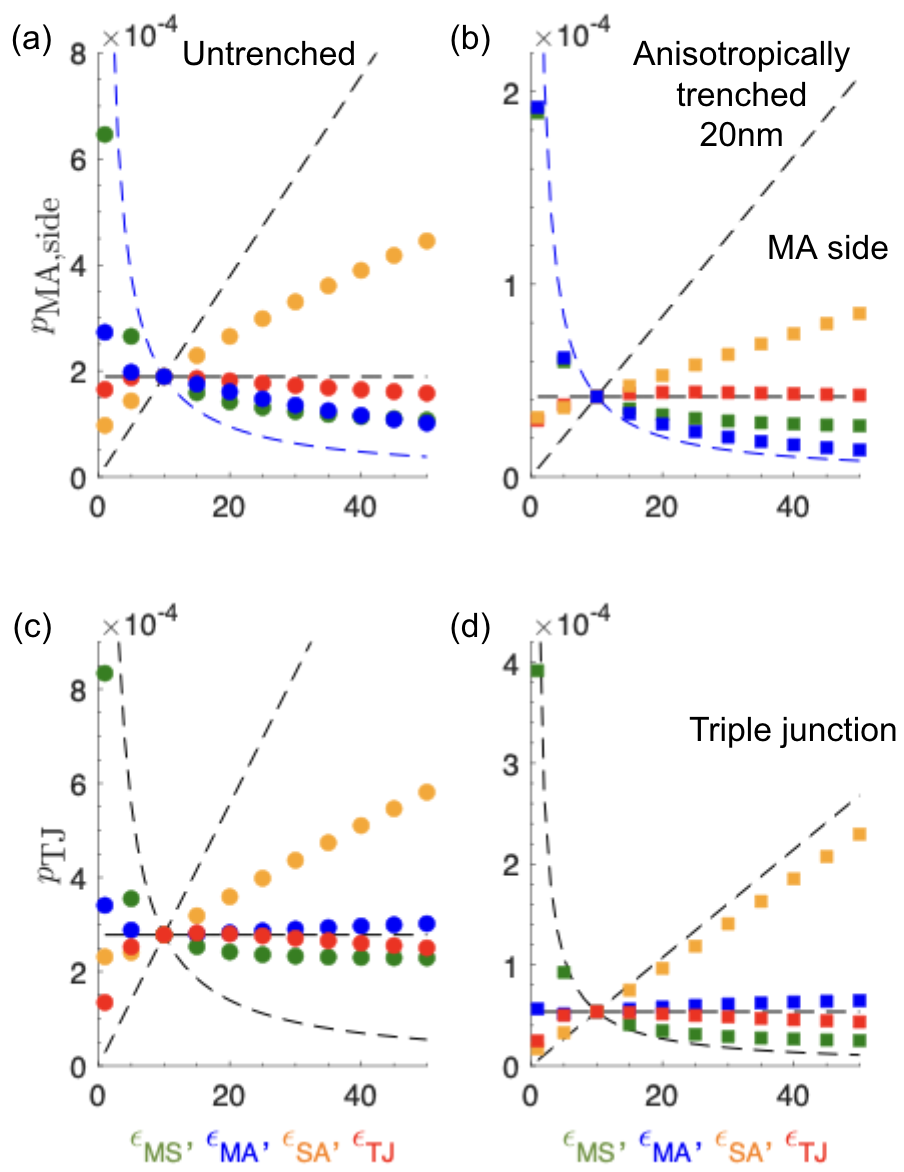}
\caption{\label{fig:dielconstlinearity}
Dielectric constant linearity analysis of interface regions for an untrenched CPW (left), and 20~nm anisotropically trenched CPW (right). 
The participation ratio $p$ is plotted as a function of each interface's dielectric constant $\varepsilon$ for (a,b) the MA side and (c,d) the TJ. The dashed lines (linear and inverse-linear) are calculated from Eq.~\ref{eq:dielconstlin} for $p_{\mathrm{sim}}$($\varepsilon_{\mathrm{sim}}$ = 10), the horizontal line indicates the simulated participation ratio $p_{\mathrm{sim}}$. Note: strong deviations from both the predicted linear and inverse‑linear trends are observed for the MA side and triple‑junction (TJ) regions, especially for the untrenched CPW.
}
\end{figure}

The MA side and TJ regions, which do not follow the expected linear relationships, are shown in Fig.~\ref{fig:dielconstlinearity}. For the untrenched cross-section, the MA side and TJ participations have very similar relationships to all dielectric constants, with a weak linear relationship to the SA dielectric constant, and a weak inversely linear relationship to MA and MS dielectric constants. This is consistent with Fig.~\ref{fig:sidewalldirection}(a) and (c). 

With 20 nm of anisotropic trenching, the MA side participation becomes less (linearly) sensitive to $\varepsilon_{\mathrm{SA}}$, more (inversely linearly) sensitive to $\varepsilon_{\mathrm{MA}}$, and insensitive to $\varepsilon_{\mathrm{TJ}}$. The TJ participation becomes more (linearly) sensitive to $\varepsilon_{\mathrm{SA}}$, more (inversely linearly) sensitive to $\varepsilon_{\mathrm{MS}}$, and insensitive to $\varepsilon_{\mathrm{MA}}$.

Our findings are consistent with the COMSOL simulations in Ref.~\citenum{Lahtinen2020}.

\section{Trenching effects and relative importance of MA sidewall}

Trenching has been shown to increase MA participation and decrease participation of other regions~\cite{Calusine2018,vissers2012reduced}. Here, we dive deeper into trenching effects on various subregions, as well as compare isotropic and anisotropic trenching as defined above. 

\begin{figure}
\includegraphics[width=0.9\linewidth]{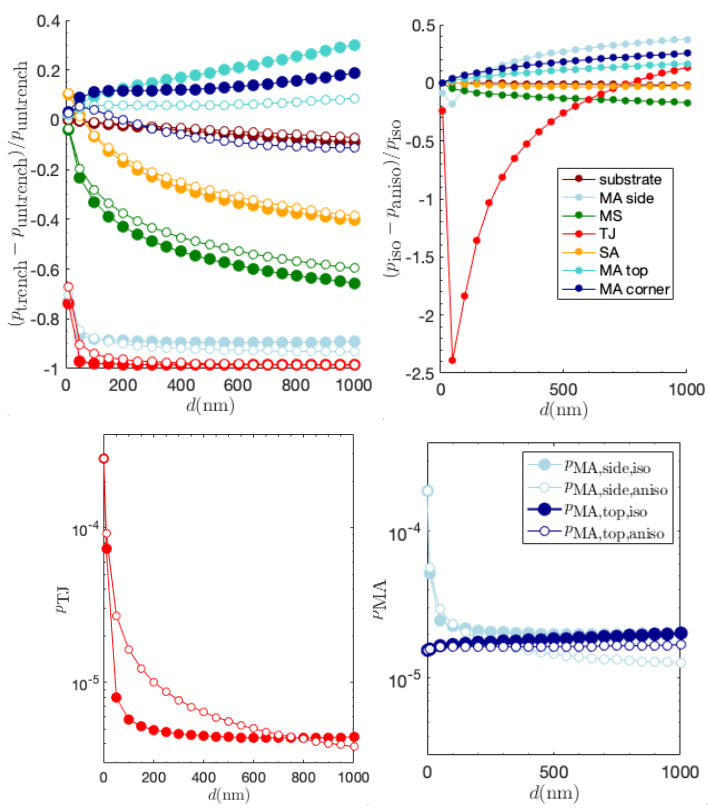}
\includegraphics[width=0.9\linewidth]{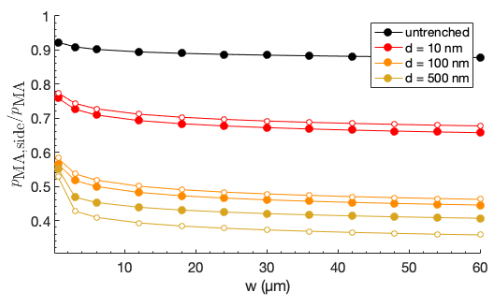}
\caption{\label{fig:maimportance}
Effects of isotropic and anisotropic trenching. (top-left) Fractional difference in participation $(p_\mathrm{trench} - p_\mathrm{untrench})/p_\mathrm{untrench}$ between trenched and untrenched CPW as a function of trench depth $d$ for both isotropic (filled circles) and anisotropic (empty circles) etching profiles. (top-right) Fractional difference in participation $(p_\mathrm{iso} - p_\mathrm{aniso})/p_\mathrm{iso}$ between devices with anisotropic and isotropic trench profiles. (middle-left) TJ interface participation as a function of trench depth $d$ for isotropic and anisotropic trenching. (middle-right) MA side participation compared to the top as a function of trench depth $d$.
(bottom) Relative participation of MA sidewall $p_{\mathrm{MA,side}}/p_{\mathrm{MA}}$ as a function of conductor width $w$. Note that gap width $g$ is also varied as $g = 0.5w$.
}
\end{figure}

Fig.~\ref{fig:maimportance}(a) shows the fractional change in participation for each region due to trenching. This allows us to identify the regions that are the most sensitive to trenching.
Trenching produces the most drastic changes in the TJ and MA sidewall regions, with the majority of the decrease in participation occurring within the first 100 nm of trenching. The MS and SA regions also see a decrease in participation, while the other regions see little change or some increase in participation.

Fig.~\ref{fig:maimportance}(b) allows us to identify the regions in which participation differs significantly between isotropic and anisotropic etch profiles. We see quite a large difference in the TJ, while the other regions do not vary more than $\sim50\%$ over 1~$\mu$m of trenching. The TJ behavior is shown in greater detail in Fig.~\ref{fig:maimportance}(c).

Recently, there has been an increased interest in superconductor capping studies, where the thin film composing the capacitor in qubits and resonators is coated with another material such as a thin film normal metal or superconductor in order to suppress MA interface loss~\cite{Bal2024,Chang2025}. It is common for only the MA top to be coated in these studies, leaving the MA side region bare. In Fig.~\ref{fig:maimportance}(d), we see that while the MA sidewall participation is higher than that of the top by an order of magnitude with no trenching, it decreases to a value roughly equal to the top after about 100 nm of trenching, and does not fall much below that with further trenching. Similar behavior for the sidewall participation was observed in the simulation of transmon qubits using finite-element methods \cite{murthy2026identifying}. Finally, we show in Fig.~\ref{fig:maimportance}(e) that unlike trenching, device size, represented here as CPW width $w$, does not significantly decrease the relative importance of the MA sidewall. Thus, we can conclude that regardless of device size, CPWs must be trenched in order to be sensitive to changes in the MA top such as capping.

\section{Conclusion}

Superconducting resonators often serve as proxies in studies of the materials and fabrication processes used for superconducting qubits. Rapid and accurate calculation of the participation ratios of thin dielectrics on the surfaces of superconducting resonators is critical to interpreting loss measurements of these devices. In this work, we demonstrate a fast boundary integral capacitance solver that is better than the standard solvers based on finite element analysis for the accurate calculation of the participation ratios of these thin regions, usually a few nanometers thick, on structures with critical dimensions three orders of magnitude larger. Its speed and accuracy enable studies where the dielectrics are divided into many more subregions or important parameters like the individual dielectric constants are swept. We highlight two examples in coplanar waveguide geometries that take advantage of these new features. The first is a careful evaluation of the linearity assumptions typically used to interpret loss measurements of these resonators. In the second, we evaluated how these thin regions, especially the sidewall of the metal near the substrate, are impacted by trenching of the substrates.

The solver is also capable of handling multiple conductors and complex geometries, e.g., flip-chip geometries with the stacked chip at a floating potential. Next steps include leveraging this solver capability to accurately calculate participation ratios of 2D slices of a 3D transmon qubit by matching the mode structure calculated by a full 3D microwave electromagnetic solver.

\begin{acknowledgments}
This work is supported in part by the Materials Characterization and Quantum Performance: Correlation and Causation (MQC) program by the Laboratory for Physical Sciences. This work is also supported in part by the U.S. Department of Energy, Office of Science, National Quantum Information Science Research Centers, Superconducting Quantum Materials and Systems Center (SQMS), under Contract Number 89243024CSC000002.

This work was conducted with the support of funding through the National Institute of Standards and Technology (NIST). Certain commercial materials and equipment are identified in this paper to foster understanding. Such identification does not imply recommendation or endorsement by NIST, nor does it imply that the materials or equipment identified is necessarily the best available for the purpose.
Thanks to internal reviewers for helpful feedback. Generative AI tools were utilized for editorial assistance, including grammar and style editing. The authors have reviewed and take full responsibility for all content.
\end{acknowledgments}

\appendix

\section{Solver Theory}

\subsection{The transmission line problem}

We consider a transmission line consisting of $N$ ideal
conductors embedded in a lossless dielectric
medium with piecewise-constant permittivity $\varepsilon(\mathbf{x})$. Transmission lines involving at least two dielectric media
cannot support a pure TEM mode, but in the low-frequency regime the fields are
nearly transversal (quasi-TEM mode), and we arrive at the electrostatic problem in two dimensions
\begin{equation}
 \nabla \cdot \varepsilon E = 0, \quad \varepsilon = \varepsilon_r \varepsilon_0, 
\end{equation}

where
$\varepsilon_0 = \frac{1}{\mu_0 c^2} \approx 
8.854$ pF/m is the permittivity of free space and $\varepsilon_r$ is the
relative permittivity.

Since the permittivity is piecewise constant and the electric field $E = -\nabla \phi$, the
scalar potential $\phi$ satisfies the boundary value problem
\begin{equation}
   \begin{split}
     \nabla \cdot \varepsilon \nabla \phi &= 0, \quad \text{outside conductors,}\\
     \phi &= V_i, \quad \text{on conductor interfaces $\Gamma_i$, \, $i=1,\dots,N$,}\\
     \phi^{+} &= \phi^{-}, \quad \text{on dielectric interfaces $\Gamma_d$,}\\
     \varepsilon^{+} \frac{\partial\phi^{+}}{\partial n} &= \varepsilon^{-} \frac{\partial\phi^{-}}{\partial n}, \quad \text{on dielectric interfaces $\Gamma_d$,}\\
     \phi \to 0 & \quad \text{as $|\xb|\to\infty$},
   \end{split} \label{extDirichlet}
\end{equation}
where $n$ is the outward normal to the interfaces, $V_i$ is the prescribed potential
on the conductor $i$, and $\phi^{\pm}$, $\varepsilon^{\pm}$ are the values of the potentials and permittivities, respectively, on the interior~($+$) and exterior~($-$) sides of dielectric interfaces.
In order to solve this exterior
boundary value problem, we represent $\phi$ as a
single layer potential
\begin{equation}
  \phi(\xb) = \int_\Gamma G(\xb,\xb') \sigma(\xb')\, d \Gamma,
\end{equation}
where $\sigma$ is an unknown potential density, $\Gamma$ is the union of conductor and dielectric interfaces, and
\begin{equation}
  G(\xb,\xb') = \frac{1}{2\pi} \log \frac{1}{|\xb-\xb'|}
\end{equation}
is the Green's function for the Laplace equation in two dimensions. It is easy to see that the scalar potential $\phi$ satisfies the Laplace equation in two dimensions; therefore, we just need to enforce the remaining boundary conditions.  The potential on each conductor is constant,
so for $\xb$ on conductor interface $\Gamma_i$
\[
\int_{\Gamma} G(\xb,\xb') \sigma(\xb')\, d \Gamma = V_i
\]
and for $\xb$ on dielectric interfaces
\begin{equation*}
    \frac{1}{2} (\varepsilon^{+} + \varepsilon^{-}) \sigma(\xb) + (\varepsilon^{+} - \varepsilon^{-}) \frac{\partial}{\partial n } \int_{\Gamma} G(\xb,\xb') \sigma(\xb')\, d \Gamma = 0.\\
\end{equation*}
Continuity of the potential across the dielectric interfaces is satisfied automatically due to continuity of the single layer representation.
The charge-neutrality condition
\begin{equation} \label{charge-neutrality}
  \sum_{i=1}^{N} Q_i = 0
\end{equation}
is needed to satisfy the condition $\phi \to 0$ at infinity, and is enforced by shifting all conductor potentials by a common offset, $V_i \to V_i - V_0$, chosen so that the resulting charges sum to zero (see Section~\ref{sec:capacitance-matrices}). Here, the net free charge on conductor $i$ is
\begin{equation} \label{netcharge}
Q_i = \int_{\Gamma_i} \varepsilon(\xb) \sigma(\xb)\,  d\Gamma_i.
\end{equation}
Note that $\sigma$ is a normalized charge density; the physical surface charge density differs by a factor of $\varepsilon$, as reflected in (\ref{netcharge}) (see also~\cite{naborswhite1992fastcap}).

\subsection{Capacitance matrices}\label{sec:capacitance-matrices}

For a system of $N$ conductors, the generalized capacitance matrix $\CCb$ gives the relationship between the conductor potentials and the net free conductor charges:
\begin{equation}
  \begin{bmatrix}
\mathcal{C}_{11} & \cdots & \mathcal{C}_{1N}\\
\vdots & \ddots & \vdots\\
\mathcal{C}_{N1} & \cdots & \mathcal{C}_{NN}
\end{bmatrix}
  \begin{bmatrix}
V_{1} \\
\vdots \\
V_{N}
\end{bmatrix} =
  \begin{bmatrix}
Q_{1} \\
\vdots \\
Q_{N}
\end{bmatrix}.
\end{equation}
To obtain column $j$ of $\boldsymbol{\mathcal C}$, we solve (\ref{extDirichlet}) using unit voltage excitations~\cite{paulfeather1976capmat}
\[
V_i^{(j)}=\delta_{ij}\quad\text{on }\Gamma_i,\qquad i=1,\dots,N,
\]
and evaluate the induced net charges $Q_i^{(j)}$.
However, the generalized capacitance matrix $\boldsymbol{\mathcal C}$ does not automatically enforce charge neutrality~(\ref{charge-neutrality}).
Following~\cite{paulfeather1976capmat,paul2007multiconductor}, a uniform potential shift $V_0$ is applied to all conductors so that the total free charge vanishes. Given prescribed potentials $\mathbf{V}$, the shift is determined by
\begin{equation} \label{V0-shift}
  V_0 = \frac{\displaystyle\sum_{m,k} \mathcal{C}_{mk}\, V_k}{\displaystyle\sum_{m,n} \mathcal{C}_{mn}}.
\end{equation}
The per-unit-length transmission-line (charge-neutral) capacitance matrix $\mathbf C$ then relates the shifted potentials to the charges. Substituting $V_i \to V_i - V_0$ into $Q_i = \sum_j \mathcal{C}_{ij} V_j$ yields
  \begin{equation}
  C_{ij} = \mathcal{C}_{ij} - \frac{\sum_m \mathcal{C}_{im} \cdot \sum_n \mathcal{C}_{nj}}{\sum_{m,n} \mathcal{C}_{mn}}.\label{capacitance0}
  \end{equation}
  By construction, $\sum_j C_{ij}=0$ and $\sum_i C_{ij}=0$, i.e., rows and
  columns of $\mathbf C$ sum to zero, reflecting charge neutrality.
  In the standard multiconductor transmission line literature~\cite{paulfeather1976capmat,paul2007multiconductor}, a reference conductor is selected and the capacitance matrix is further reduced by eliminating the corresponding row and column. We retain the full $N \times N$ matrix $\mathbf C$, since the choice of reference conductor can be ambiguous in multi-conductor geometries.
The total electric field energy (per unit length) for a system of $N$ conductors is given by the quadratic form
\begin{equation}\label{energy-quadratic}
    W = \frac{1}{2} \mathbf{V}^T \mathbf{C} \, \mathbf{V},
\end{equation}
where $\mathbf{V}$ is the vector of conductor potentials.
For a two-conductor system, (\ref{energy-quadratic}) reduces to
$W = \frac{1}{2} C V^2$, where $C = C_{11}$ and $V = V_2 - V_1$.

\subsection{Participation ratios}

Recall that the participation ratio for the domain $\Omega_i$ is
defined as
  \[
    p_i = \frac{W_i}{W},
    \]
    where  $W_i$ is the energy per unit length in domain $\Omega_i$:
    \[
    W_i = \int_{\Omega_i} \frac{\varepsilon_i}{2} |E|^2\,dS,
  \]
  and $W$ is the total electric field energy per unit length:
  \[  W = \sum_i W_i, \]
  $dS$ is the area element, $\varepsilon_i$ is the permittivity of domain $\Omega_i$.
We compute the participation ratios using two methods. The first method uses Green's first identity~\cite{strauss2008partial} to express the domain integral as a boundary integral:
\[
   p_i = W^{-1}  \int_{\Omega_i} \frac{\varepsilon_i}{2}  |\nabla \phi|^2\,dS =  W^{-1} \int_{\partial \Omega_i} \frac{\varepsilon_i}{2} \phi \, \frac{\partial \phi}{\partial n}\,ds,
\]
where $ds$ is the arc length element and $n$ is the outward-pointing (exterior) unit normal.
This allows us to recover the participation ratios using a single unperturbed solution,
by integrating the product of the normal component of the $E$ field and
the scalar potential on interfaces adjacent to the domain of interest.
Since the scalar potential is bounded, the singularities to be integrated are
dominated by the behavior of the normal component of the $E$ field.

The second method uses perturbation analysis~\cite{schneider1969microstrip,spielman1977dissipation}:
\[
   p_i = \frac{\varepsilon_i}{W} \frac{\partial W}{\partial \varepsilon_i}.
\]
This requires multiple solves with perturbed dielectric constants to evaluate the numerical derivatives of the energy functional~$W$. To avoid the numerical instabilities of finite-difference differentiation with small perturbations, we use complex step differentiation~\cite{Martins2003}, complexifying $\varepsilon_i$ with a small imaginary perturbation~$h$:
\[
  \frac{\partial W}{\partial \varepsilon_i}
      \approx \operatorname{Im} \frac{ W(\ldots, \varepsilon_i + \sqrt{-1}\, h, \ldots)}{h}.
\]
In practice, we use $h = 10^{-12}$ and treat the perturbative result as an independent validation of the participation ratios computed via Green's first identity.
This requires the solver to handle complex-valued permittivities and capacitances, which is straightforward in the BIE framework.

\section{Solver Implementation Details}

This appendix provides technical implementation details of the boundary integral equation solver used to generate the results in this paper. The solver is implemented in MATLAB/Octave with Fortran MEX acceleration for computationally intensive operations.

\subsection{Geometry discretization}

The solver divides conductor and dielectric interface boundaries into segments (referred to as chunks), where both geometry $\mathbf{r}(t) = (x(t), y(t))$ and surface densities $\sigma(t)$ are represented using order $p$ Legendre polynomial expansions on each chunk's parameter domain $t \in [0,1]$. The geometry discretization provides boundary points, normal vectors, and arc length quadrature weights $w_i$ for discretizing the boundary integral operators. Gauss-Legendre nodes (typically 10-16 per chunk) are used for evaluation, providing spectral accuracy. 

Adaptive refinement is achieved through either uniform refinement or dyadic
refinement with independent left/right/middle refinement control for
handling geometric singularities at conductor and dielectric interface
corners and junctions.

\subsection{Boundary integral operators}

The single-layer potential operator $S_0$ and its normal derivative $S_0'$ form the basis of the integral equation formulation. For a potential density $\sigma$ on boundary $\Gamma$,
\begin{align}
  S_0[\sigma](\xb) &= \int_\Gamma G(\xb,\xb') \sigma(\xb')\, d\Gamma, \\
  S_0'[\sigma](\xb) &= \frac{\partial}{\partial n} \int_\Gamma G(\xb,\xb') \sigma(\xb')\, d\Gamma.
\end{align}

For conductors, assuming that the normal $n$ points outward from the conductor, the discretized system matrix uses
\begin{equation}
  \mathcal{A}_{\text{cond}} = S_0 + S_0' + \frac{1}{2}\,I,
\end{equation}
 where $I$ is the identity matrix and the combined operator forms a second-kind Fredholm integral equation, ensuring the discretized system matrix is well-conditioned.
 The null space of $S_0' + \frac{1}{2}\,I$, 
 corresponding to charge distributions with zero field
inside the conductors, is removed by $S_0$, which yields
  a unique solution~\cite{rachh2016integral}.
  An alternative formulation using $S_0$ alone is available as a solver         
  option, but provides much slower convergence.

For dielectric interfaces, the jump conditions across the boundary yield
\begin{equation}
  \mathcal{A}_{\text{diel}} = \frac{\varepsilon^{+}-\varepsilon^{-}}{\varepsilon^{+}+\varepsilon^{-}} S_0' + \frac{1}{2} I,
\end{equation}
where $\varepsilon^{+}$ and $\varepsilon^{-}$ are the permittivities on the positive (interior) and negative (exterior) sides of the interface, respectively, following the convention from (\ref{extDirichlet}).

The discretization of the boundary integral equations for all conductor and dielectric interface boundaries results in a linear system
\begin{equation}\label{discretizedSystem}
  \mathbf{A} \boldsymbol{\xi} = \mathbf{b},
\end{equation}
where $\mathbf{A}$ is the assembled system matrix with blocks obtained by discretizing $\mathcal{A}_{\text{cond}}$ for conductor boundaries and $\mathcal{A}_{\text{diel}}$ for dielectric interface boundaries, $\boldsymbol{\xi}$ is the vector of unknown charge densities at all discretization nodes, and $\mathbf{b}$ encodes the boundary conditions from (\ref{extDirichlet}).

\subsection{Matrix discretization}

The boundary integral equations are discretized using a Galerkin-type scheme that is converted to Nyström form for efficient preconditioning~\cite{GREENGARD2021100092,Askham_chunkIE_a_MATLAB_2024,bremer2012nystrom}.
On each chunk, the potential density $\sigma$ is represented as a Legendre polynomial expansion of order $p$, sampled at Gauss-Legendre nodes. For each pair of boundary segments, the kernel $K$ of $\mathcal{A}_{\text{cond}}$ or $\mathcal{A}_{\text{diel}}$ is first integrated over the source chunk at auxiliary target points $\xb_m = \xb(t_m)$, where $\{t_m\}$ are nodes on the target parameter domain chosen to integrate both smooth functions and endpoint logarithmic singularities:
\begin{equation}
  I_{mn} = \int_{\Gamma_j} P_n(s)\, K(\xb_m, \yb)\, d\ell,
\end{equation}
where $P_n$ is the $n$-th Legendre polynomial on the source parameter domain. The source integration uses oversampled Gauss-Legendre quadrature for well-separated chunks, adaptive recursive subdivision for nearby chunks, and generalized Gaussian quadrature (designed to handle the logarithmic singularity in~$S_0$) for self-interactions. The results are then projected onto the target chunk's Legendre basis:
\begin{equation}
  (\mathbf{A}_{ij})_{\ell n} = \sum_{m=1}^{n_{\text{aux}}} w_m\, P_\ell(t_m)\, I_{mn},
\end{equation}
where $w_m$ are the quadrature weights associated with the nodes $\{t_m\}$ and $(\mathbf{A}_{ij})_{\ell n}$ are the resulting Galerkin matrix elements with respect to the Legendre basis on the source and target chunks. The discretization is converted from Legendre coefficients to values at Gauss-Legendre nodes, yielding a node-to-node formulation that facilitates square-root-weight preconditioning.

\subsection{Iterative solver and fast multipole acceleration}

The linear system (\ref{discretizedSystem}) is solved using the generalized minimal residual (GMRES) iterative method.
Square-root-weight preconditioning is applied by default to improve convergence, particularly for highly refined meshes. The system is transformed as
\begin{equation}
  \mathbf{W}^{1/2} \mathbf{A} \mathbf{W}^{-1/2} \yb = \mathbf{W}^{1/2} \mathbf{b}, \quad \mathbf{W} = \text{diag}(\mathbf{w}),
\end{equation}
where $\yb = \mathbf{W}^{1/2} \boldsymbol{\xi}$ and $\mathbf{w}$ are the discretization quadrature weights~\cite{bremer2012nystrom}. This ensures the standard inner product in the preconditioned system corresponds to the $L_2$ inner product on the boundary, so that GMRES minimizes the residual in the physically meaningful $L_2$ norm.

For large problems, the solver employs the Fast Multipole Method (FMM)~\cite{greengard1987fmm,gimbutas2015fmm} to reduce the computational complexity of applying the system matrix. The dense system matrix is never formed; matrix-vector products are computed on-the-fly using FMM for far-field interactions (segment separation-to-size ratio $> 2.4$) and precomputed direct evaluation for near-field interactions (ratio $\leq 2.4$). This locally corrected approach enables efficient iterative solution, as near-field corrections are computed once and reused at each GMRES step~\cite{GREENGARD2021100092}.

\subsection{Capacitance and participation ratio computation}

The generalized capacitance matrix $\boldsymbol{\mathcal{C}}$ is assembled from $N$ linear solves (one per unit-voltage excitation, Appendix~\ref{sec:capacitance-matrices}), and the physical capacitance matrix $\mathbf{C}$ follows from~(\ref{capacitance0}).
The participation ratios are then computed using Green's first identity. For region $\Omega_i$ with boundary $\partial\Omega_i$,
\begin{equation}
  p_i = W^{-1}  \int_{\partial \Omega_i} \frac{\varepsilon_i}{2}\,\phi \, \frac{\partial \phi}{\partial n} \, ds,
\end{equation}
where $W$ is the total electric field energy per unit length corresponding to the scalar potential $\phi$.
For a two-conductor geometry, the charge-neutrality condition $Q_1 + Q_2 = 0$ with unit voltage difference $V_2 - V_1 = 1$ uniquely determines the quasi-TEM mode. The charge density $\sigma$ is obtained as a linear combination of the two unit-voltage solutions, with coefficients chosen to satisfy these constraints.
Given $\sigma$, the potential $\phi = S_0[\sigma]$ and the interior normal derivative $\partial\phi/\partial n$ from $S_0'[\sigma]$ are evaluated on each region boundary, and the boundary integral above yields~$p_i$. The deviation of $\sum_i p_i$ from unity provides an internal consistency check.

For geometries with more than two conductors, the physical quasi-TEM modes are identified via the eigenvalue problem~\cite{paulfeather1976capmat,paul2007multiconductor}
\begin{equation}
  \mathbf{L}_0 \mathbf{C}\,\mathbf{v}_m = \mu_0 \varepsilon_0 \, \varepsilon_{\mathrm{eff},m}\,\mathbf{v}_m,
\end{equation}
where $\mathbf{L}_0 = \mu_0 \varepsilon_0 \, \mathbf{C}_0^{+}$ is the external inductance matrix and $\mathbf{C}_0^{+}$ is the pseudoinverse of the free-space capacitance matrix (obtained by setting all relative permittivities to one). The single spurious eigenvalue ($\varepsilon_\mathrm{eff}=0$) is discarded. The unit-voltage solutions are each adjusted by a uniform potential shift~(\ref{V0-shift}) to enforce charge neutrality~(\ref{charge-neutrality}), and the mode charge density $\sigma_m$ is the linear combination of these adjusted solutions with coefficients $\mathbf{v}_m$. The participation ratios are computed as in the two-conductor case using the mode energy $W_m = \frac{1}{2}\,\mathbf{v}_m^T \mathbf{C}\,\mathbf{v}_m$.

\subsection{Software implementation}

The solver is implemented as a hybrid MATLAB/Octave and Fortran codebase, leveraging the strengths of each language. MATLAB/Octave handles high-level logic, geometry generation, iterative linear solvers (GMRES), pre- and post-processing, and visualization. Fortran MEX extensions accelerate computationally intensive operations, including matrix assembly, FMM evaluation, Legendre polynomial expansions, and specialized quadrature routines.

The implementation follows similar design principles to chunkIE~\cite{Askham_chunkIE_a_MATLAB_2024}, a MATLAB integral equation toolbox for chunk-based boundary integral equation discretization, and FMMLIB2D~\cite{gimbutas2015fmm}, a Fortran library for 2D fast multipole method kernels. The Fortran-MATLAB interface is generated using MWrap~\cite{Bindel_mwrap_2011}, a MEX gateway generator that automatically creates the necessary wrapper code.

\section{Exact Solution Comparisons\label{Exact}}

\subsection{Two circular wires}

For two parallel circular wires with radii $r_1$ and $r_2$ separated by center-to-center distance $D$, the exact capacitance per unit length is given by~\cite{smythe1968static}
\begin{equation}
  C = \frac{2\pi\varepsilon_0\varepsilon_r}{\cosh^{-1}\left(\frac{D^2 - r_1^2 - r_2^2}{2r_1 r_2}\right)}.
\end{equation}

We investigate convergence for the configuration with wire radii $r_1 = \qty{0.45}{\micro\metre}$ and $r_2 = \qty{0.25}{\micro\metre}$, separated by center-to-center distance $D = \qty{1.5}{\micro\metre}$ in a medium with $\varepsilon_r = 1.0$. The exact capacitance is $C_{\text{exact}} = 19.403122346044$ pF/m (using CODATA(2014) vacuum permittivity value). Table~\ref{tab:twowire_convergence} shows numerical results for chunk refinement at fixed polynomial order $p = 10$.

\begin{table}[h]
\caption{\label{tab:twowire_convergence}Convergence of two-wire transmission line capacitance with chunk refinement at fixed polynomial order $p=10$. The exact capacitance is $C_{\text{exact}} = 19.403122346044$ pF/m.}
\begin{ruledtabular}
\begin{tabular}{cccc}
$n_{\text{chunks}}$ & $N_{\text{pts}}$ & $C$ (pF/m) & Rel. Error \\
\hline
4  & 40  & 19.403120211097 & $1.1 \times 10^{-7}$ \\
8  & 80  & 19.403122343569 & $1.3 \times 10^{-10}$ \\
12  & 120 & 19.403122345997 & $2.4 \times 10^{-12}$ \\
16  & 160 & 19.403122346038 & $2.9 \times 10^{-13}$ \\
\end{tabular}
\end{ruledtabular}
\end{table}

\subsection{Square coaxial line}

For concentric square conductors with inner square half-width $a$ (side length $2a$) and outer square half-width $b$ (side length $2b$), the exact capacitance per unit length is obtained using the Anderson method~\cite{anderson1950capacitance}. The solution is expressed in terms of the complete elliptic integral of the first kind $K(k)$, where $k$ is the elliptic modulus, and the complementary elliptic integral, defined as $K'(k) = K(\sqrt{1-k^2})$.
First, solve the transcendental equation
\begin{equation}
  \frac{K(p)}{K'(p)} = \frac{1 - a/b}{1 + a/b}
\end{equation}
for the modulus $p$.

Then compute the complementary modulus $q = \sqrt{1-p^2}$ and a new elliptic modulus $x = p^2/q^2$.
The capacitance per unit length is given by
\begin{equation}
  C = 2\varepsilon_0\varepsilon_r\frac{K'(x)}{K(x)}.
\end{equation}

Note that in numerical implementations, MATLAB's \texttt{ellipke(m)} uses the parameter $m = k^2$, so to compute $K(k)$ one calls \texttt{ellipke($k^2$)}.

We investigate convergence for the configuration with $a =
  \qty{0.25}{\micro\metre}$, $b = \qty{1}{\micro\metre}$, and $\varepsilon_r
   = 1.0$, which has the exact capacitance $C_{\text{exact}} =42.893400549082$
  pF/m (using CODATA(2014) vacuum permittivity value). At fixed polynomial order $p = 10$, we compare uniform chunk
  refinement (Table~\ref{tab:uniform_square_coax_convergence}) with dyadic chunk refinement (Table~\ref{tab:dyadic_square_coax_convergence}).
  We observe that dyadic chunk refinement, which concentrates discretization
   points near corners, achieves comparable
  accuracy with significantly fewer total discretization points than uniform refinement. Comparison of the square coaxial waveguide results obtained from our solver and Ansys Maxwell simulations is given in Table~\ref{tab:squarecoax_comparison}.
  
\begin{table}[h]
\caption{\label{tab:uniform_square_coax_convergence}Convergence of square coaxial line capacitance with uniform chunk refinement at fixed polynomial order $p=10$. The exact capacitance is $C_{\text{exact}} = 42.893400549082$ pF/m.}
\begin{ruledtabular}
\begin{tabular}{cccc}
$n_{\text{chunks}}$ & $N_{\text{pts}}$ & $C$ (pF/m) & Rel. Error \\
\hline
32  & 320  & 42.89272337912 & $1.6 \times 10^{-5}$ \\
128  & 1280 & 42.89329399506 & $2.5 \times 10^{-6}$ \\
512  & 5120 & 42.89338377156 & $3.9 \times 10^{-7} $ \\
2048  & 20480 & 42.89339790693   & $6.2 \times 10^{-8} $ \\
8192  & 81920 & 42.89340013298   & $9.8 \times 10^{-9} $ \\
\hline
32768 & 327680 & 42.89340048354   & $1.5 \times 10^{-9} $ \\
131072 & 1310720 &  42.89340053873  & $2.4 \times 10^{-10} $ \\
\end{tabular}
\end{ruledtabular}
\end{table}

\begin{table}[h]
\caption{\label{tab:dyadic_square_coax_convergence}Convergence of square coaxial line capacitance with dyadic chunk refinement at fixed polynomial order $p=10$. The exact capacitance is $C_{\text{exact}} = 42.893400549082$ pF/m.}
\begin{ruledtabular}
\begin{tabular}{cccc}
$n_{\text{chunks}}$ &  $N_{\text{pts}}$ & $C$ (pF/m) & Rel. Error \\
\hline
32  & 320  & 42.89272337912 & $1.6 \times 10^{-5}$ \\
48 & 480  & 42.89313197114 & $6.3 \times 10^{-6}$ \\
80 & 800 & 42.89335826928 & $9.9 \times 10^{-7}$ \\
112 & 1120 & 42.89339389116 & $1.5 \times 10^{-7}$ \\
144 & 1440 & 42.89339950055 & $2.4 \times 10^{-8}$ \\
\hline
208 & 2080 & 42.89340052306 & $6.1 \times 10^{-10}$ \\
272 & 2720 & 42.89340054841 & $1.5 \times 10^{-11}$ \\
\end{tabular}
\end{ruledtabular}
\end{table}

\begin{table}[htp!]
\caption{\label{tab:squarecoax_comparison}Comparison of the square coaxial waveguide results obtained from our solver and Ansys Maxwell simulations.}
\begin{ruledtabular}
\begin{tabular}{c|cc}
 & This work & Ansys Maxwell \\
\hline
Capacitance (pF/m)  & 42.8934 & 42.893 \\
\hline
log10(energy density [J/m$^3$]) & \includegraphics[width=0.13\textwidth]{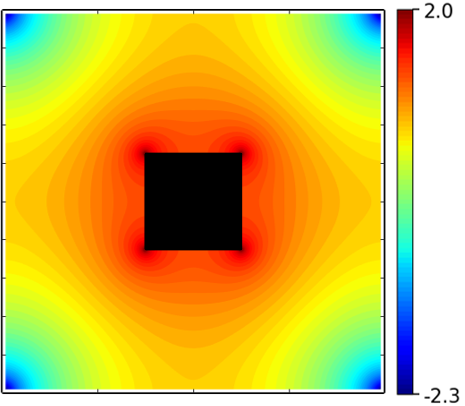} & \includegraphics[width=0.13\textwidth]{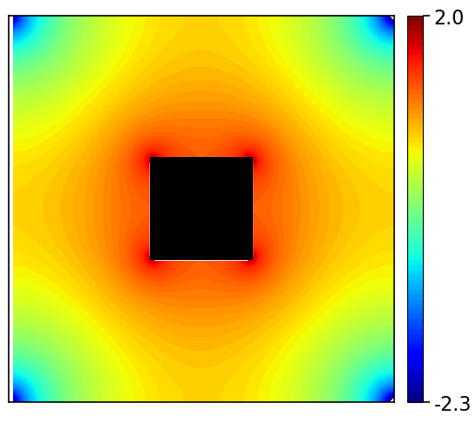} \\
\hline
Potential (V) & \includegraphics[width=0.12\textwidth]{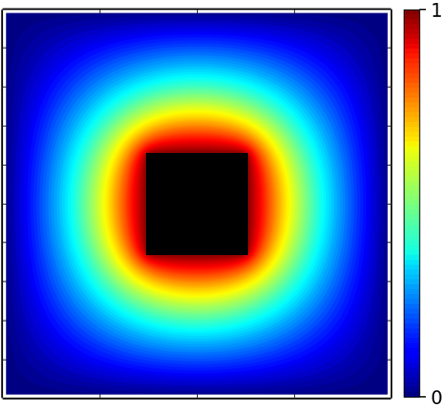}& \includegraphics[width=0.12\textwidth]{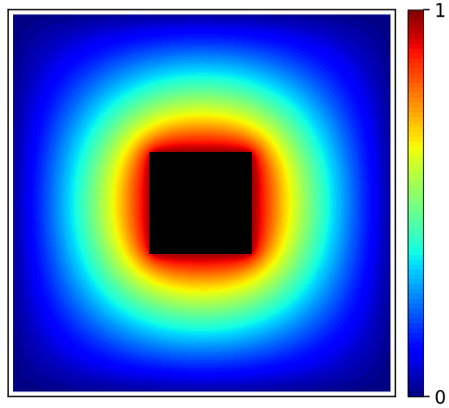} \\
\end{tabular}
\end{ruledtabular}
\end{table}

\section{Comparison to FEM solvers}

In this section, we provide detailed comparisons between our solver and two commonly-used FEM electrostatic solvers: Ansys Maxwell and COMSOL.

\subsection{Comparison to Ansys Maxwell}

For the Ansys Maxwell comparison, we use a geometry with $w$ = 6~$\mu$m, $g$ = 3~$\mu$m, $h$ = 100~nm, and trench depth $d$ = 0~nm, and we vary ground plane width $f$. All dielectric constants are set to 10 for simplicity.

The boundary conditions for our solver are most similar to the balloon boundary condition used in Maxwell, so this is the option used in this comparison. These boundaries emulate an open environment in which the charge at infinity balances the charge within the simulation region, resulting in zero net charge. Neumann boundary conditions can lead to a discrepancy in the simulated capacitance and participation values. In the future, a symmetry boundary condition could be introduced to our solver. This would further reduce the domain size of symmetric geometries such as idealized CPW cross-sections, and thus run time, by a factor of two.

To determine the computational resources required for accurate results, we perform a systematic convergence study of our solver, and compare this to Ansys Maxwell. Total capacitance is a convenient metric for identifying convergence, and we use it here along with participation ratios.

\begin{figure}
\includegraphics[width=1.0\linewidth]{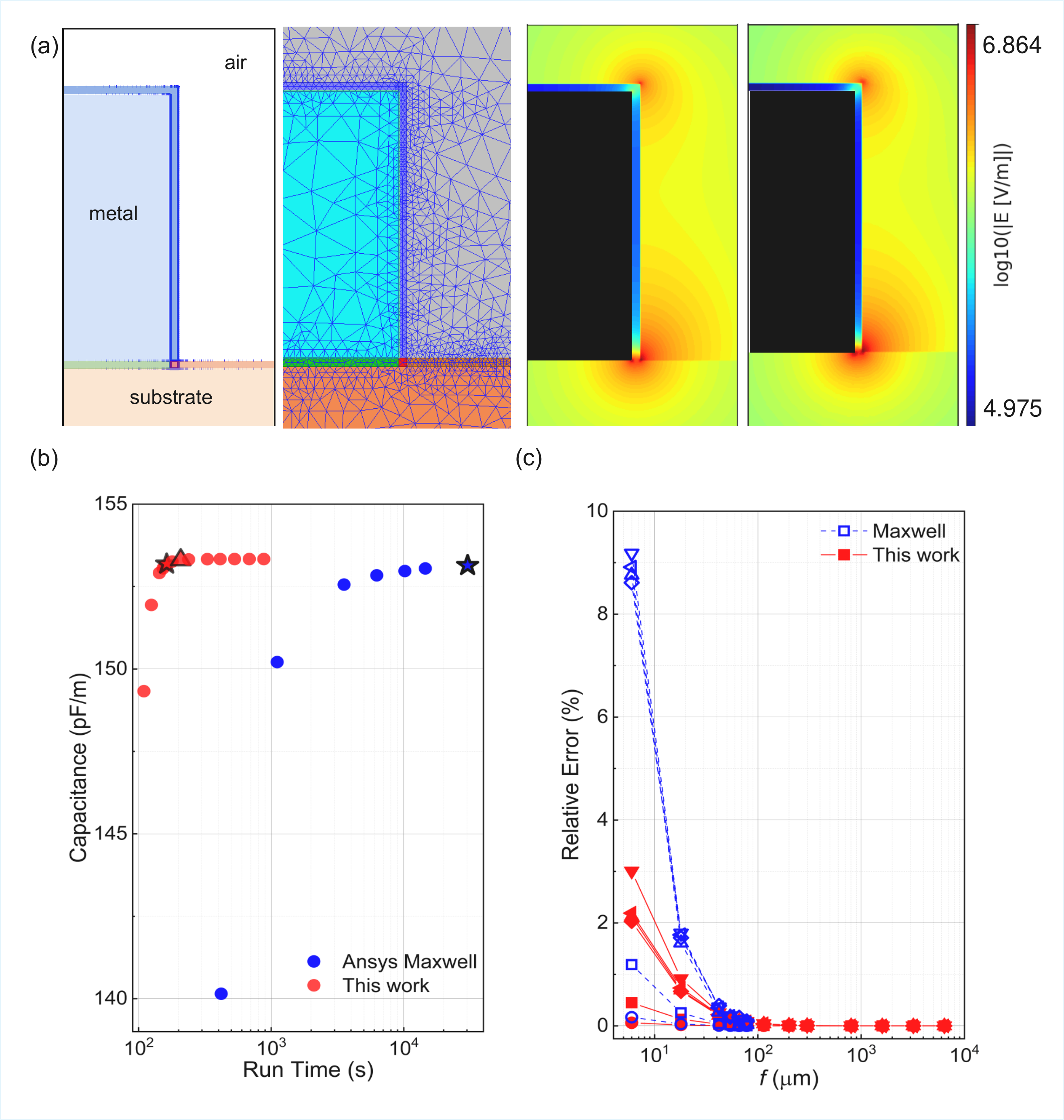}
\caption{CPW convergence analysis and comparison to Ansys Maxwell. (a) Cross-sectional visualization of the CPW geometry. The left pair shows the chunk discretization (left) from our solver and the mesh (right) from Ansys Maxwell. The right pair shows the corresponding simulated electric-field distributions, plotted as the normalized field magnitude $\log_{10}(|E|/E_{\max})$ for $f$ = 114~$\mu$m cases (denoted with stars in (b)). (b) Convergence of the total capacitance as a function of solver run time, where increased run time corresponds to increasing the ground-plane width $f$. Star markers denote $f = 114~\mu$m and triangle markers denote $f = 300~\mu$m. (c) Relative error of the computed participation ratios as a function of $f$. Blue dashed curves (empty markers) correspond to Ansys Maxwell and red solid curves (solid markers) correspond to our solver. Marker shapes indicate different dielectric interfaces: squares--air; circles--substrate; upward triangles--MA; downward triangles--MS; diamonds--TJ; and left-pointing triangles--SA.}
\label{fig:convergence}%
\end{figure}

Discretization for both solvers is shown in Fig.~\ref{fig:convergence}(a). Meshing for our solver is performed using the discretization best practices outlined above. In Maxwell, the mesh density is refined near interfaces and high-field regions, with a maximum element size of 1~nm in interface regions.

A significant factor determining the accuracy and precision of a capacitance solution is the domain size. As explained above, our solver implements an infinite domain size in the +y and -y directions. In Maxwell, we find that the total capacitance quickly converges to the asymptotic limit when increasing y domain size (see Appendix~D for details). Here, we set the Maxwell y-direction domain size to 150~$\mu$m, corresponding to 75~$\mu$m of air above and 75~$\mu$m of substrate below the CPW.

We now vary the x-direction domain size, defined by ground plane width $f$, in order to identify convergence of total capacitance $C$. All simulations are performed on a six-core Intel workstation. We expect $C$ to increase to an asymptotic limit as we increase the x-direction domain size and include more previously-truncated capacitive regions, and this behavior is seen in Fig.~\ref{fig:convergence}(b).

Table~\ref{tab:maxwellcomp_v4} shows a detailed breakdown of participation values from the simulations denoted as stars in Fig.~\ref{fig:convergence}(b). All participation ratios from our solver are within 1~$\%$ of those from the Ansys Maxwell simulation except for $p_{\text{MA,corner}}$, and our solver is faster by a factor of approximately 170. The largest discrepancy between the two solvers is the participation of the $p_{\text{MA,corner}}$ region.

We define relative error as the fractional difference between a solution and the solution from the longest-running simulation. As shown in Fig.~\ref{fig:convergence}(c), relative errors below 1~$\%$ for all participation ratios are obtained for $f > 20$~$\mu$m for our solver, and $f > 40$~$\mu$m for Ansys Maxwell. Note that this indicates that participation ratios will be significantly perturbed in experiment if other on-chip components are at a distance less than 20-40~$\mu$m from the CPW.

\begin{table}[tbp!]
\caption{\label{tab:maxwellcomp_v4} Capacitance and participation ratio comparison for $f$ = 114~$\mu$m. These results are derived from the simulations denoted as stars in Fig.~\ref{fig:convergence}. } 
\begin{ruledtabular}
\begin{tabular}{c|ccc}
& \multicolumn{3}{c}{Untrenched ($d$ = 0~nm)} \tabularnewline
 & Maxwell & This work & Difference\\
\hline
$t$ (min) & 506 & 3 & 170x \\
\hline
$C$ (pF/m) & 153.15 & 153.26 & 0.07$\%$ \\
$p_{\text{MA,top}}$ (ppm) & 15.6 & 15.5 & 0.3$\%$ \\
$p_{\text{MA,side}}$ (ppm) & 188.6 & 189.2 & 0.3$\%$ \\
$p_{\text{MA,corner}}$ (ppm) & 5.2 & 5.1 & 3$\%$ \\
$p_{\text{TJ}}$ (ppm) & 281 & 278 & 0.9$\%$ \\
$p_{\text{SA}}$ (ppm) & 1740 & 1736 & 0.2$\%$ \\
$p_{\text{MS}}$ (ppm) & 1699 & 1698 & 0.02$\%$ \\
$p_{\text{sub}}$ (ppm) & 899547 & 899428 & 0.01$\%$ \\
$p_{\text{air}}$ (ppm) & 96524 & 96650 & 0.1$\%$ \\
\hline
& \multicolumn{3}{c}{Isotropically trenched ($d=$100 nm)} \tabularnewline
 & Maxwell & This work & Difference\\
 \hline
 $t$ (min) & 549 & 3.2 & 171x \\
\hline
$C$ (pF/m) & 145.66 & 145.47 & 0.1$\%$ \\
$p_{\text{MA,top}}$ (ppm) & 17.13 & 17.06 & 0.4$\%$ \\
$p_{\text{MA,side}}$ (ppm) & 22.8 & 22.5 & 1$\%$ \\
$p_{\text{MA,corner}}$ (ppm) & 5.8 & 5.6 & 3$\%$ \\
$p_{\text{TJ}}$ (ppm) & 6.1 & 5.6 & 8$\%$ \\
$p_{\text{SA}}$ (ppm) & 1615 & 1602 & 0.8$\%$ \\
$p_{\text{MS}}$ (ppm) & 1136 & 1123 & 1$\%$ \\
$p_{\text{sub}}$ (ppm) & 887953 & 887434 & 0.05$\%$ \\
$p_{\text{air}}$ (ppm) & 109245 & 109791 & 0.4$\%$ \\
\end{tabular}
\end{ruledtabular}
\end{table}

\subsection{Neumann boundary conditions in Ansys Maxwell}
To evaluate the influence of boundary conditions in Ansys Maxwell, we simulate the untrenched CPW geometry using Neumann boundary conditions on all outer boundaries of the computational domain. Although Neumann boundaries are often used to approximate electric-field decay in electrostatic problems, they enforce a vanishing normal displacement field $D$ at the truncation surface and therefore do not accurately represent an open environment. As a result, the CPW fringing fields interact artificially with the truncation boundary, leading to deviations in both the total capacitance and the electric field distribution, even though these deviations are small as shown in Table~\ref{tab:bcscomp_v1}. Here we summarize the comparison between balloon and Neumann boundaries for the same geometry $f$ = 114~$\mu$m, and also the extracted values using our solver.

\begin{table}[h]
\caption{\label{tab:bcscomp_v1} Capacitance and participation ratio comparison for $f$ = 114~$\mu$m. These results are derived from the simulations in Maxwell with different boundary conditions and this work as well.} 
\begin{ruledtabular}
\begin{tabular}{c|ccc}
 & Balloon & Neumann & This work \\
 \hline
$C$ (pF/m) & 153.15 & 153.13 & 153.26 \\
$p_{\text{MA,top}}$ (ppm) & 15.590 & 15.589 & 15.545 \\
$p_{\text{MA,side}}$ (ppm) & 188.596 & 188.593 & 189.247 \\
$p_{\text{MA,corner}}$ (ppm) & 5.2318 & 5.2315 & 5.0659 \\
$p_{\text{TJ}}$ (ppm) &280.756 & 280.752 & 278.216 \\
$p_{\text{SA}}$ (ppm) & 1739.838 & 1739.832 & 1736.081 \\
$p_{\text{MS}}$ (ppm) & 1698.89 & 1698.87 & 1698.46 \\
$p_{\text{sub}}$ (ppm) & 899547 & 899568 & 899428 \\
$p_{\text{air}}$ (ppm) & 96524 & 96503 & 96650 \\
\end{tabular}
\end{ruledtabular}
\end{table}

\subsection{Ansys Maxwell y-axis truncation convergence}
To ensure a consistent comparison between our boundary-integral formulation solver and Ansys Maxwell, we performed a convergence study on the y-direction truncation size used in the Maxwell simulations. Because Maxwell requires a finite simulation region, the computational domain was constructed as a rectangular box with a total height extending above the air region and below into the substrate. The truncation height in the y-direction was swept from 10~$\mu$m to 150~$\mu$m, and the extracted capacitance was found to converge at a truncation height of 100~$\mu$m. In the main text, we choose 150~$\mu$m for Ansys Maxwell y-direction truncation to obtain more accurate values for extracted capacitance and participation ratios.  

\begin{figure}[htp!]
\includegraphics[width=0.8\linewidth]{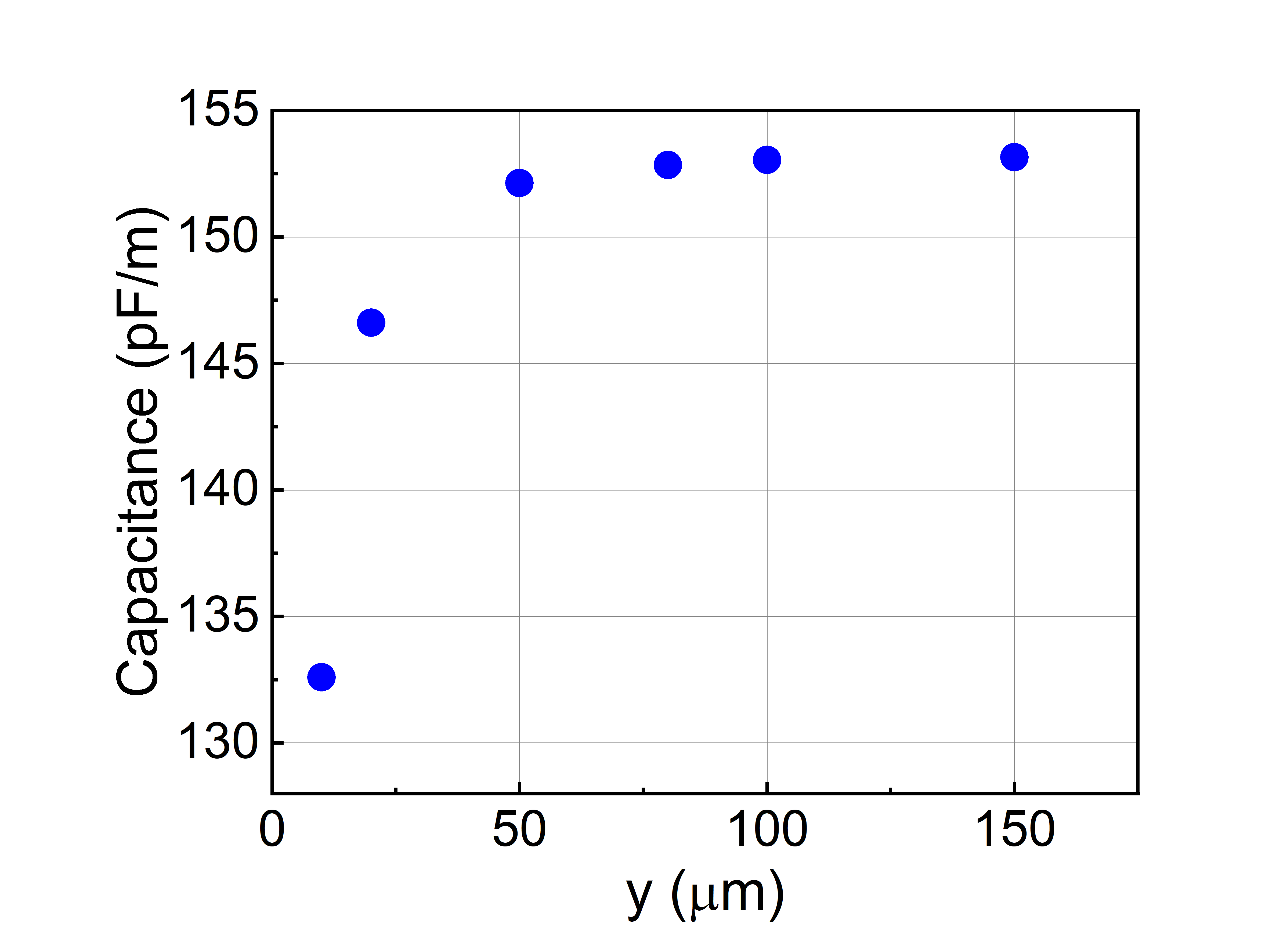}
\caption{
Capacitance versus y-direction truncation for the Ansys Maxwell simulation. 
\label{fig:capacitancery}} %
\end{figure}

\subsection{Capacitance relative error plots}
\begin{figure}[htp!]
\includegraphics[width=0.9\linewidth]{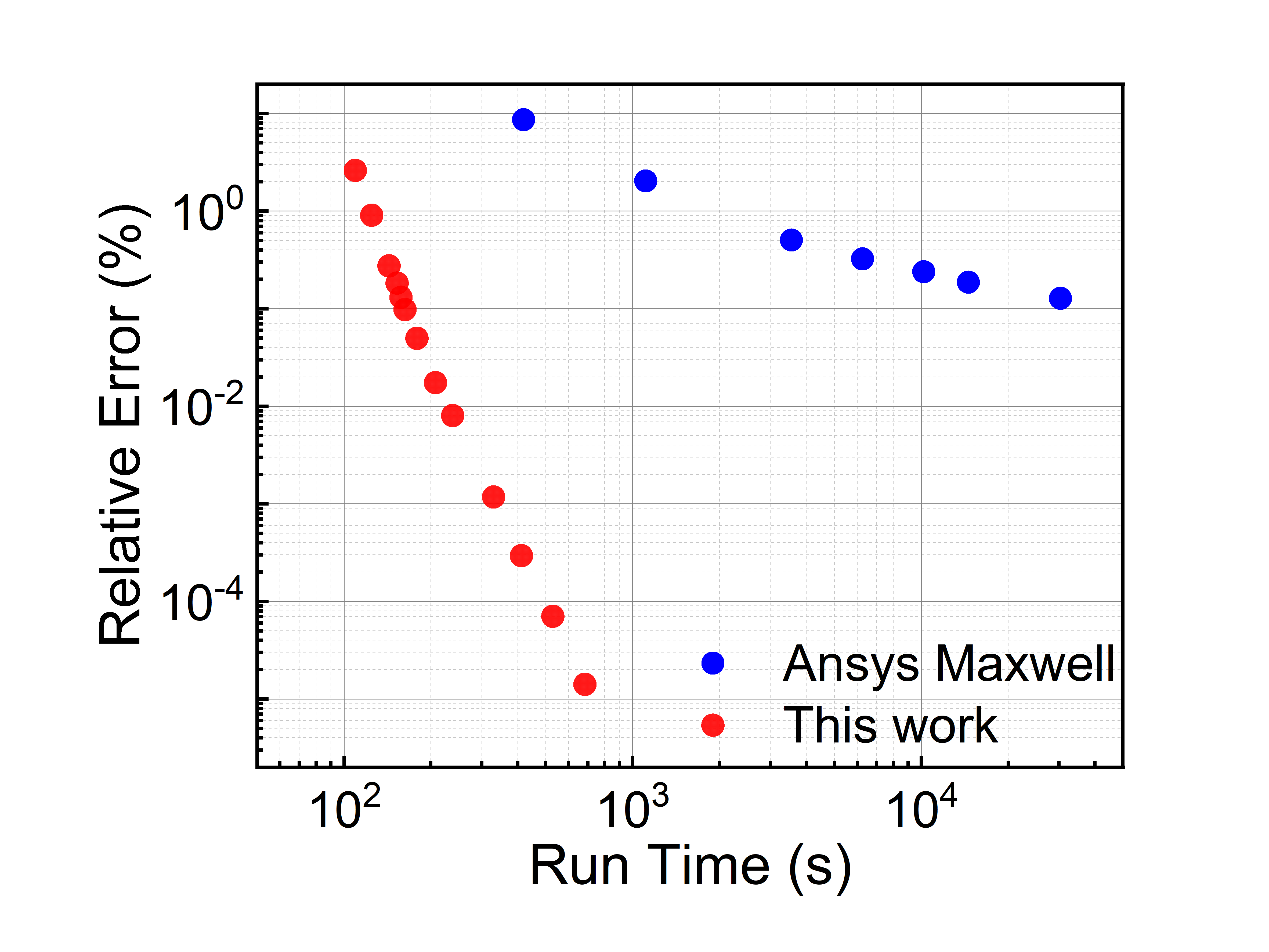}
\includegraphics[width=0.9\linewidth]{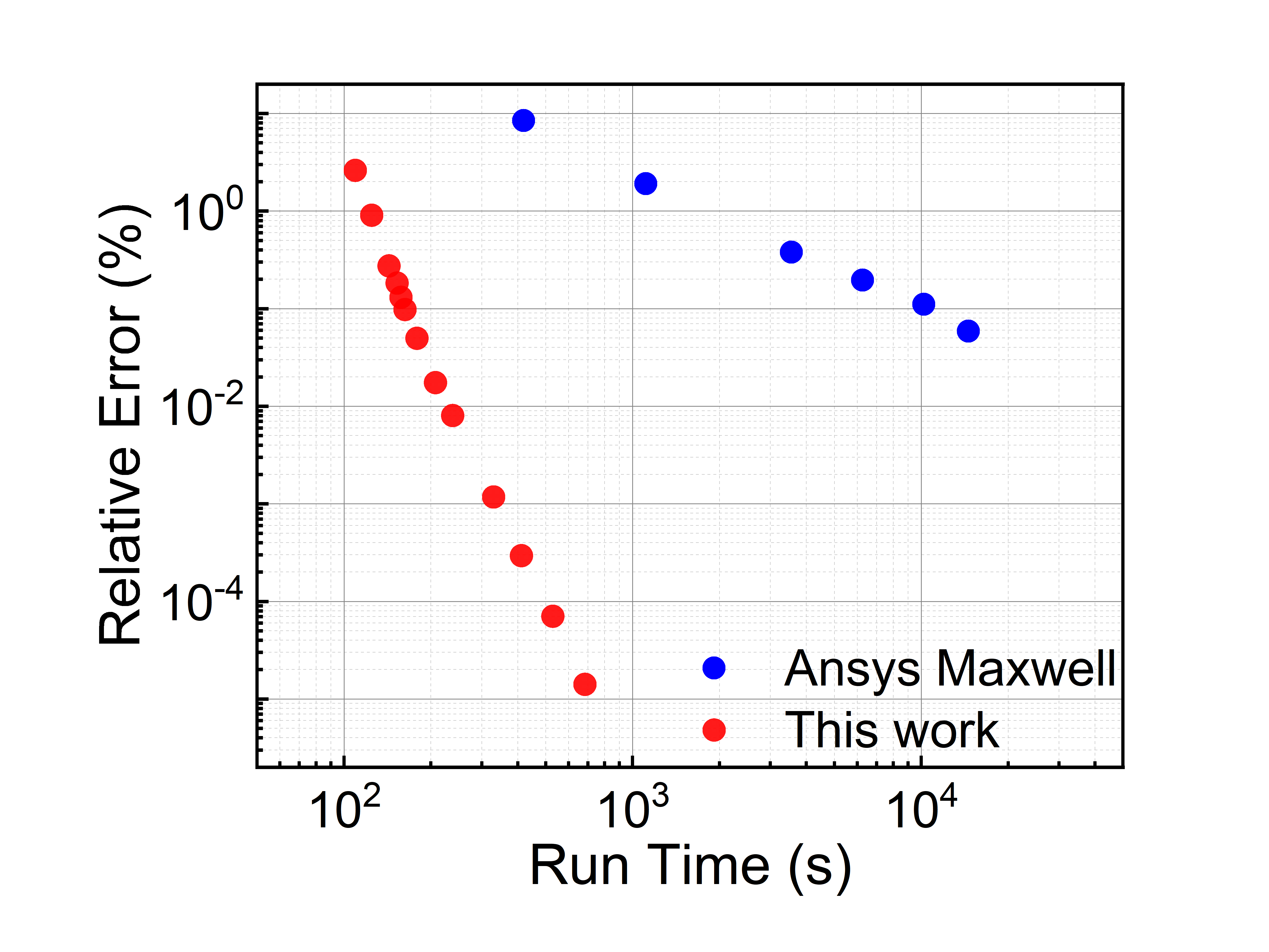}
\caption{
Top: Relative error versus runtime for the simulation configurations, with the relative error defined with respect to the final most converged value of our solver.  
Bottom: Relative error versus runtime for the simulation configurations, with the relative error defined separately using the final converged value for each method as the reference. 
\label{fig:errorruntime}} %
\end{figure}

To quantify the convergence behavior, we compute the relative error in the extracted capacitance as a function of runtime. Fig.~\ref{fig:errorruntime}(a) shows the relative error when all methods are referenced to the final most converged value obtained from our solver. This provides a unified view of absolute convergence across solvers. Fig.~\ref{fig:errorruntime}(b) shows the relative error defined separately for each method using the method’s own most converged value as the reference, which highlights the rate of convergence of each solver. These plots demonstrate that our method attains high accuracy orders of magnitude faster than the alternatives.

\subsection{Comparison to COMSOL}

We compare our results to the COMSOL simulations in Wenner et al.~\cite{Wenner2011} using the same untrenched CPW geometry as in that work. For their example c3[a], with parameters $w$ = 5~$\mu$m, $g$ = 2~$\mu$m, $h$ = 0.1~$\mu$m, $d$ = 0~$\mu$m, $dt$ = 3~nm, our solver (using the ground plane width $f$ = 300~$\mu$m)
yields the capacitance of $C$ = 163.8~pF/m, compared to Wenner's result of 163~pF/m. The participation ratios show similarly excellent agreement, reproducing previously published values obtained from more time-intensive simulations to within fractions of a percent (Table~\ref{tab:wennercomparison}).

\begin{table}[h]
\caption{\label{tab:wennercomparison} Wenner et al.~\cite{Wenner2011} comparison. Example c3[a].}
\begin{ruledtabular}
\begin{tabular}{cccc}
 & Wenner & This work & Difference \\
\hline
$C$ (pF/m) & 163 & 163.8  &  0.49$\%$\\
$p_{\text{MA}}$ (ppm) & 290  & 290.1 & 0.03$\%$\\
$p_{\text{TJ}}$ (ppm) & 387  & 386.1 &  0.23$\%$\\
$p_{\text{SA}}$ (ppm) & 2286 & 2283 & 0.13$\%$\\
$p_{\text{MS}}$ (ppm) & 2234 & 2231 &  0.13$\%$\\
\end{tabular}
\end{ruledtabular}
\end{table}

\section{Dielectric linearity of all regions}

\begin{figure*}[th!]
\includegraphics[width=0.4\linewidth]{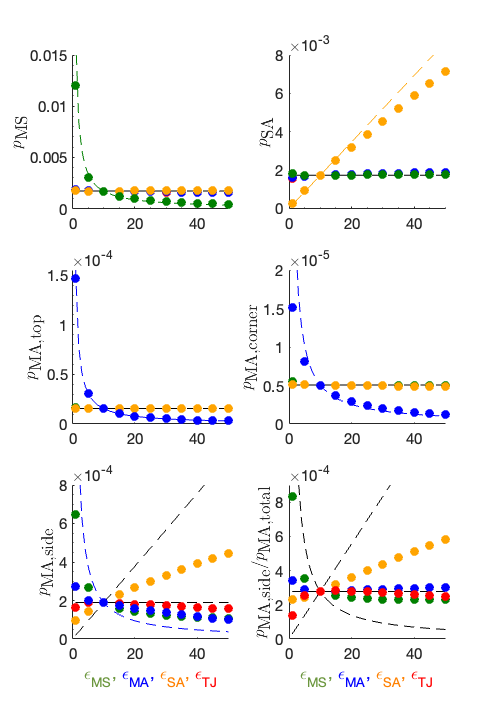}
\includegraphics[width=0.4\linewidth]{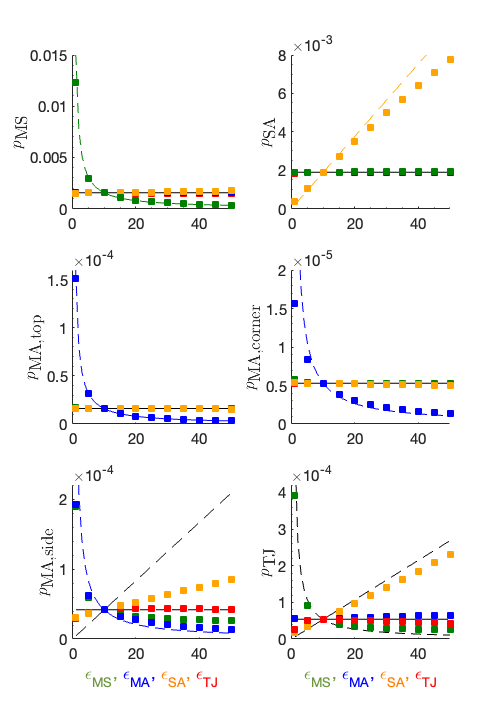}
\caption{\label{fig:dielconstlinearity_suppl_untrench}
Dielectric constant linearity analysis of interface regions for an untrenched (left) and trenched (right) CPW. In each pannel, the participation ratio $p$ is plotted as a function of each interface's dielectric constant $\varepsilon$ for MS (top-left), SA (top-right), MA top (middle-left), MA corner (middle-right), MA side (bottom-left), and TJ (bottom-right). The dashed lines (linear and inverse-linear) are calculated from Eq.~\ref{eq:dielconstlin} for $p_{\mathrm{sim}}$($\varepsilon_{\mathrm{sim}}$ = 10); the horizontal line shows to the participation ratio $p_{\mathrm{sim}}$. Note: a pronounced deviation from the inverse‑linear relationship is observed for the MA side and triple‑junction (TJ) regions, especially in the untrenched CPW.
}
\end{figure*}

The dielectric linearity for every region is shown in Fig.~\ref{fig:dielconstlinearity_suppl_untrench}.

\bibliography{mainbib}

\end{document}